# Large Fermi Surface of Heavy Electrons at the Border of Mott Insulating State in NiS$_2$

S. Friedemann,[1][*][†] H. Chang,[2][†] M. B. Gamża,[3,4] P. Reiss,[2] X. Chen,[2] P. Alireza,[2] W. A. Coniglio,[5] D. Graf,[5] S. Tozer,[5] and F. M. Grosche[2][*]

[1] HH Wills Laboratory, University of Bristol, Bristol, BS8 1TL, UK.

[2] Cavendish Laboratory, University of Cambridge, Cambridge, CB3 0HE, UK.

[3] Department of Physics, Royal Holloway, University of London, Egham TW20 0EX, UK.

[4] Jeremiah Horrocks Institute for Mathematics, Physics and Astrophysics, University of Central Lancashire, Preston PR1 2HE, UK.

[5] National High Magnetic Field Laboratory, Tallahassee, FL 83810, U.S.A.

[†]These authors contributed equally to this work

**One of the early triumphs of quantum physics is the explanation why some materials are metallic whereas others are insulating. While a treatment based on single electron states correctly predicts the character of most materials this approach can fail spectacularly, when the electrostatic repulsion between electrons causes strong correlations. Not only can these favor new and subtle forms of order in metals, such as magnetism or superconductivity, they can even cause the electrons in a half-filled energy band to lock into position altogether, producing a correlated, or Mott insulator (*1*, *2*). Arguably the most extreme manifestation of electronic correlations in dense electronic matter, the transition into the Mott insulating state raises a number of fundamental questions. Foremost among these is the fate of the electronic Fermi surface and the associated charge carrier mass, as the Mott transition is approached at low temperature, which have been particularly controversial in the high temperature superconducting cuprates. We report the first direct observation of the Fermi surface on the metallic side of a Mott insulating transition by high pressure quantum oscillatory measurements in NiS$_2$. Our results point at a large Fermi surface consistent with Luttinger's theorem and a strongly enhanced carrier effective mass, suggesting that electron localization occurs via a diverging effective mass and concomitant slowing down of charge carriers as predicted theoretically (*3*).**

**Introduction**

Prototypical Mott insulators such as $V_2O_3$, NiS, NiO, and $NiS_2$ feature singly occupied $d$ states with reduced orbital overlap and strong on-site Coulomb repulsion $U$ (*2*). The electronic states which would make up a band straddling the chemical potential $\mu$ if $U$ were zero, split at large $U$ into a lower and upper Hubbard band, falling on either side of the chemical potential (*4*). This gaps out charged excitations and produces an insulating ground state (as sketched in the inset of Fig. 1.). The electronic spectrum can be tuned, for instance, by varying the lattice density and thereby the underlying electronic bandwidth $W$ under applied pressure, making it possible to close the charge gap and metallize the Mott insulator. The resulting correlated metal displays a narrow peak in the density of states at the chemical potential associated with long-lived Fermi-liquid quasi-particles (*5*, *6*).

Luttinger's theorem (*7*) fixes the volume enclosed by the Fermi surface in this correlated metallic state as identical to that of the corresponding uncorrelated metal. With the Fermi surface volume unaffected by proximity to the MIT, the insulating state is realized, within the canonical Brinkman-Rice picture (*3*), via a suppression of the quasiparticle weight $Z$ and, consequently, a divergence of the effective mass $m^*$.

Despite the long history of the Brinkman-Rice picture of a MIT, there is surprisingly little direct experimental evidence for its key predictions – the large Fermi surface expected for half-filled bands, and the enhanced effective mass on the threshold of the Mott state. This can be attributed in part to practical difficulties: accessing the threshold of the Mott state requires finely controlled metallization of a Mott insulator either by pressure or chemical substitution. Pressure precludes the use of angular resolved photoemission spectroscopy (ARPES), whereas the disorder associated with substitution seriously hinders high-resolution Fermi surface measurements by quantum oscillatory techniques. Quantum oscillation measurements have been adapted to high pressures in the past but are particularly demanding for MIT, because the supremely important quality of the sample in the metallic state can only be assessed in the metallic phase at high pressure, slowing down the screening process. ARPES, on the other hand, can suffer from complications arising from conducting surface states (*8*) and requires exquisite energy resolution to disentangle coherent and incoherent contributions to the spectrum, to determine conclusively the shape of the Fermi surface in 3D and to track accurately the evolution of $m^*$.

Detailed quantum oscillation studies near a MIT have so far mostly been conducted in two material systems – the superconducting cuprates and the 2D organic charge transfer salts. In the cuprates, however, charge density wave reconstruction, superconductivity with high critical fields and the pseudogap obscure the direct access to the Fermi surface emerging from the doped Mott state (*9*). Transport, and optical spectroscopy show indications for a small Fermi surface with no change in effective mass in underdoped cuprates (*10*). The mass enhancement found in quantum oscillation studies of cuprate superconductors have been attributed to putative quantum critical point(s) beneath the superconducting dome(*11*), unrelated to the threshold of the Mott state. In 2D organic conductors quantum oscillation studies have been limited to samples which are already metallic at ambient pressure (*12*), preventing access to the immediate vicinity of the MIT.

**Results**

We study the 3D inorganic Mott insulator $NiS_2$, which is free from superconductivity and charge density wave reconstruction. $NiS_2$ can be tuned into the metallic state by hydrostatic pressure, enabling immediate access to the Fermi surface geometry and quasiparticle mass in close proximity to the MIT. We employ quantum oscillation measurements to focus on the coherent quasiparticle states which lie at the heart of the Brinkman-Rice framework. This approach contrasts with spectroscopic measurements on the Ni(S/Se)$_2$ composition series, which give important and comprehensive insight on the summed coherent and incoherent contributions to the spectral function as the MIT is crossed by chemical substitution (*13*–*16*). It complements recent ARPES results which are interpreted in terms of a progressive reduction of the

Fermi velocity on approaching the MIT (*17*), but by focusing on the coherent part of the spectrum offers definitive and high resolution measurements of the Fermi surface and effective mass in a 3D material.

The pyrite $NiS_2$ has long been identified as a prototypical Mott insulator alongside NiS and NiO (*2*, *18–25*). The sulphur atoms in $NiS_2$ form dimers, yielding a valence of 2 for the Ni atoms like in NiO and NiS (*25*, *26*). The Ni *d* states are split by the Coulomb interaction into a lower and upper Hubbard band with band edges at -1 eV and 3.5 eV below and above the chemical potential, respectively (*25*). Sulphur dimers contribute an antibonding $p^*_\sigma$ band 1eV above the chemical potential. Under pressure, the sulphur dimers are rigid but the inter-dimer hopping increases, causing the $p^*_\sigma$ band to broaden and eventually to connect with the lower Hubbard band, inducing metallization. Whereas megabar pressures are required to metallize NiO, rendering quantum oscillation measurements impossible with current techniques, the tunability of the sulfur dimer-derived bands in $NiS_2$ pushes down the metallization pressure to about 3 GPa, where the key impediment to high pressure quantum oscillation measurements – pressure inhomogeneity caused by the pressure medium – is well understood and under control.

Sulphur vacancies are known to compromise crystal quality in $NiS_2$. Employing a Te-flux growth technique (*27*) we obtained crystals with ultra-low sulphur vacancy concentration (cf. Supplementary information I). Our high-quality single crystals of $NiS_2$ display clear insulating behavior at low pressure *p* (Fig. 1). We observe the magnetic transition into the weak ferromagnetic state at $T_{WF}$ = 30 K in agreement with previous results (*28*) (cf. Supplementary information II).

The application of hydrostatic pressure in our liquid-medium patterned-anvil cell (cf. Materials and Methods) first yields very little change at low temperature *T* (Fig. 1). A drastic change is observed at a pressure of 3 GPa: Here, the low-temperature resistivity is reduced by many orders of magnitude, signaling the MIT. The metallic state below the MIT temperature $T_{MIT}$ is recognized from the positive slope of the resistivity $d\rho/dT>0$. Increasing the pressure beyond 3 GPa results in a further reduction of the low-temperature resistivity and a shift of $T_{MIT}$ to higher temperatures.

Based on the resistivity measurements we can construct the *T-p* phase diagram for our $NiS_2$ samples (Fig. 2). We find $T_{WF}$ and $T_{MIT}$ very similar to previous high pressure transport studies (*28*, *29*). The slightly higher pressure scale is attributed to the absence of sulphur vacancies in our samples (cf. Supplementary information I). We followed the MIT up to 4.3 GPa and 185 K. The transition becomes less distinct towards the critical end point of the first order MIT line around 200 K (*29*).

The abrupt decrease of the residual resistivity by 6 orders of magnitude on crossing the MIT by applied pressure is highlighted in the inset of Fig. 2. This is followed by a further decrease of more than 2 orders of magnitude in the metallic phase, indicating the correlated nature of the material in proximity to the Mott-MIT. Importantly, with $\rho_0$ falling below 1μΩ cm, high-resolution Fermi surface studies with quantum oscillation measurements become possible.

We measure quantum oscillations in the resistivity by a contactless tunnel-diode-oscillator (TDO) method for increased resolution and to circumvent difficulties with making contacts to the sample. A single crystal $NiS_2$ sample is placed inside the micro-coil mounted in the hole of a gasket between two opposing moissanite anvils (*30*) (see Materials and Methods). Fig. 3(A) displays the quantum oscillation signal observed at a pressure of 3.8 GPa, where the residual resistivity is well below 1 μΩ cm. We find clear oscillations at high fields beyond 25 T. The Fourier transform in Fig. 3(B) identifies the oscillation frequency to be 6 kT.

The temperature dependence of the quantum oscillation amplitude can be used to extract the quasiparticle mass. This is illustrated for two subsequent runs in Fig. 4. Best fits to the data yield effective masses of $m^*=5(1)\ m_e$ and $m^*=7(2)\ m_e$, respectively. Combining the two datasets we conclude the quasiparticle mass to be $m^*=6(2)\ m_e$. Despite the uncertainty in $m^*$, our measurement provides high significance for the comparison with band structure calculations, to which we now turn.

The electronic structure of metallic $NiS_2$ has been calculated within the generalized gradient approximation, neglecting strong correlations. We use the simple cubic crystal lattice with lattice constants estimated for our crystals at 3.8 GPa (Supplementary information III). At the pressures studied, $NiS_2$ is suggested to be antiferromagnetic at low temperature (*29*). High pressure diffraction studies have so far not resolved the spin structure in metallic $NiS_2$ (*31*, *32*), but neutron diffraction studies in the Ni(S/Se)$_2$ composition series (e.g. (*33*)) suggest that the antiferromagnetic ordering wave vectors in the metallic state are (1 0 0) and its symmetry-related equivalents, as in ambient pressure, insulating $NiS_2$ at intermediate temperatures. In the latter case, a type 1 antiferromagnetic structure has been suggested (*34*), for which the magnetic unit cell is the same as the structural unit cell and no Fermi surface reconstruction would necessarily be expected. Our calculation predicts a metallic ground state, in good agreement with previously published local density approximation results (*25*, *35–37*). Modelling the insulating state would require including onsite Coulomb repulsion in the spirit of the Mott-Hubbard model (*38*). Here, we compare calculated metallic Fermi surface with our measurements in the metallic state at high pressure. The major Fermi surface sheets obtained in the band structure calculations are presented in Fig. 5: they consist of a large "Cube"-like pocket in the center of the Brillouin zone, a network connected along the edges of the Brillouin zone, and several closed pockets in the corner (only the biggest of which is presented in Fig. 5). From the predicted Fermi surface we can identify the most likely orbit to produce quantum oscillations: The small curvature on the "Cube" makes it a strong candidate. Indeed, the predicted frequency of 6.1 kT for the belly orbit (highlighted in Fig. 5) shows excellent agreement with the observed quantum oscillation frequency of 6 kT (cf. Tab. 1). In particular, it is the only orbit large enough to yield a frequency above 3 kT (cf. Supplementary information III).

**Discussion**
Whereas the predicted and observed frequencies are in excellent agreement we find a strong deviation between the predicted and observed effective mass (Tab. 1). The strong mass enhancement by nearly an order of magnitude observed in high pressure $NiS_2$ can be attributed to strong correlations present in proximity to the Mott-MIT, as would be expected within the Brinkman-Rice picture (*3*). This mass enhancement should be accompanied by a similarly large enhancement of the Sommerfeld coefficient of the specific heat capacity $\gamma$ over the band structure value. A roughly four-fold increase of $\gamma$ has indeed been observed in the composition series Ni(S$_{1-x}$Se$_x$)$_2$ on approaching the Mott transition from the metallic side (*39*, *40*). The mass enhancement is consistent, also, with the narrow band features observed in ARPES studies on metallic members of the Ni(S$_{1-x}$Se$_x$)$_2$ series (*13*). Together with the cross-sectional area determined from the 6kT orbit and assuming a spherical Fermi surface geometry, the measured mass corresponds to a Fermi velocity $v_F \sim 0.4$ eV Å/$\hbar$, comparable to results from ARPES measurements at the corresponding doping of x=0.5 (*17*).

High quality crystals of $NiS_2$ have enabled a fresh look at the Mott insulator transition in this prototypical material using a novel combination of high pressure measurement techniques. Our quantum oscillation measurements unambiguously show that coherent quasiparticles with a large Fermi surface and significant mass enhancement exist near the border of Mott localization in pressure metallized $NiS_2$. These findings offer a direct view on the emergence of a heavy Fermi liquid on the metallic side of the Mott transition and motivate further, detailed high pressure quantum oscillation studies in $NiS_2$ and other Mott insulators.

**Materials and Methods**
**Single crystals** of $NiS_2$ were grown from Te-flux as described earlier (*27*). Sulphur occupation was determined from single-crystal x-ray diffraction as detailed in Supplementary information I and are found to be in very good agreement with estimates based on the temperature dependence of the resistivity (*22*).

**Resistivity studies** were conducted in an alumina Bridgman anvil cell with patterned anvils (*41*) with Flourinert as pressure transmitting medium. The superconducting transition temperature of a lead samples was used to determine the pressure at low temperatures.

**Quantum oscillation measurements** were measured with the contactless tunnel diode oscillator (TDO) technique, which is a parallel inductor-capacitor oscillator sustained by a tunnel diode. The inductor in the experimental setup is a five turn cylindrical coil with an inner diameter of 200 μm wound with a 15 μm insulated wire. For high pressure measurement, the coil is positioned in a 300 μm diameter hole in a BeCu gasket mounted between the anvils of a Moissanite anvil pressure cell. An oriented single crystal sample of $NiS_2$ is placed in the coil. Changes in the sample resistivity in its metallic state are detected as proportional changes in the TDO resonance frequency.

A 4:1 mixture of methanol-ethanol is used as pressure transmitting medium for highly hydrostatic conditions. Sample pressure at low temperature is determined via ruby fluorescence. Quantum oscillation measurements were carried out at the NHMFL Tallahassee, with the TDO oscillating at ~250MHz, in a top-loading $^3$He cryostat in magnetic fields up to 31T.

# References

1. N. F. Mott, The Basis of the Electron Theory of Metals, with Special Reference to the Transition Metals. *Proc. Phys. Soc. Sect. A*. **62**, 416–422 (1949).

2. M. Imada, A. Fujimori, Y. Tokura, Metal-insulator transitions. *Rev. Mod. Phys.* **70**, 1039–1263 (1998).

3. W. F. Brinkman, T. M. Rice, Application of Gutzwiller's Variational Method to the Metal-Insulator Transition. *Phys. Rev. B*. **2**, 4302–4304 (1970).

4. J. Hubbard, Electron Correlations in Narrow Energy Bands. *Proc. R. Soc. A Math. Phys. Eng. Sci.* **276**, 238–257 (1963).

5. A. Fujimori *et al.*, Evolution of the spectral function in Mott-Hubbard systems with d 1 configuration. *Phys. Rev. Lett.* **69**, 1796–1799 (1992).

6. A. Georges, G. Kotliar, Hubbard model in infinite dimensions. *Phys. Rev. B*. **45**, 6479–6483 (1992).

7. J. Luttinger, Fermi Surface and Some Simple Equilibrium Properties of a System of Interacting Fermions. *Phys. Rev.* **119**, 1153–1163 (1960).

8. D. D. Sarma *et al.*, Metal-insulator crossover behavior at the surface of NiS 2. *Phys. Rev. B*. **67**, 155112 (2003).

9. N. Doiron-Leyraud *et al.*, Quantum oscillations and the Fermi surface in an underdoped high-Tc superconductor. *Nature*. **447**, 565–568 (2007).

10. W. J. Padilla *et al.*, Constant effective mass across the phase diagram of high-$T_{c}$ cuprates. *Phys. Rev. B*. **72**, 060511 (2005).

11. B. J. Ramshaw *et al.*, Quasiparticle mass enhancement approaching optimal doping in a high-Tc superconductor. *Science*. **348**, 317–320 (2015).

12. J. Caulfield *et al.*, Magnetotransport studies of the organic superconductor kappa -(BEDT-TTF) 2 Cu(NCS) 2 under pressure: the relationship between carrier effective mass and critical temperature. *J. Phys. Condens. Matter*. **6**, 2911–2924 (1994).

13. A. Matsuura *et al.*, Electronic structure and the metal-insulator transition in NiS2-xSex. *Phys. Rev. B*. **53**, R7584–R7587 (1996).

14. A. Y. Matsuura *et al.*, Metal-insulator transition in NiS 2 − x Se x and the local impurity self-consistent approximation model. *Phys. Rev. B*. **58**, 3690–3696 (1998).

15. A. Fujimori *et al.*, Resonant photoemission study of pyrite-type NiS2, CoS2 and FeS2. *Phys. Rev. B*. **54**, 16329–16332 (1996).

16. K. Mamiya *et al.*, Photoemission study of the metal-insulator transition in NiS2-xSex. *Phys. Rev. B*. **58**, 9611–9614 (1998).

17. H. C. Xu *et al.*, Direct Observation of the Bandwidth Control Mott Transition in the NiS$_{2-x}$Se$_x$ Multiband System. *Phys. Rev. Lett.* **112**, 087603 (2014).

18. J. A. Wilson, Charge Density Waves: Sliding and Related Phenomena in NbSeFormula and Other Transition-Metal Chalcogenides. *Philos. Trans. R. Soc. A Math. Phys. Eng. Sci.* **314**, 159–177 (1985).

19. J. M. Honig, J. Spalek, Electronic Properties of NiS2-xSex Single Crystals: From Magnetic Mott-Hubbard Insulators to Normal Metals. *Chem. Mater.* **10**, 2910–2929 (1998).

20. J. A. Wilson, G. D. Pitt, Metal-insulator transition in NiS2. *Philos. Mag.* **23**, 1297–1310 (1971).

21. T. Miyadai *et al.*, Neutron Diffraction Study of NiS$_2$ with Pyrite Structure. *J. Phys. Soc. Japan.* **38**, 115–121 (1975).

22. G. Krill *et al.*, Electronic and magnetic properties of the pyrite-structure compound NiS2 : influence of vacancies and copper impurities. *J. Phys. C Solid State Phys.* **9**, 761 (1976).

23. D. W. Bullett, Electronic structure of 3d pyrite- and marcasite-type sulphides. *J. Phys. C Solid State Phys.* **15**, 6163–6174 (1982).

24. R. Kautz, M. Dresselhaus, D. Adler, A. Linz, Electrical and Optical Properties of NiS$_2$. *Phys. Rev. B.* **6**, 2078–2082 (1972).

25. J. Kuneš *et al.*, Metal-insulator transition in NiS$_{2-x}$Se$_x$. *Phys. Rev. B.* **81**, 035122 (2010).

26. N. Kawai, S. Mochizuki, Insulator-metal transition in NiO. *Solid State Commun.* **9**, 1393–1395 (1971).

27. X. Yao, J. M. Honig, Growth of nickel dichalcogenides crystals with pyrite structure from tellurium melts [NiS2, NiS2-xSex (x <= 0.7)]. *Mater. Res. Bull.* **29**, 709–716 (1994).

28. N. Mori, T. Watanabe, Pressure effects on the magnetic transition temperatures of NiS2. *Solid State Commun.* **27**, 567–569 (1978).

29. N. Takeshita *et al.*, Quantum criticality and disorder in the antiferromagnetic critical point of NiS$_2$ pyrite. *arXiv:0704.0591v1 [cond-mat.str-el]* (2007) (available at http://arxiv.org/abs/0704.0591).

30. P. L. Alireza, S. R. Julian, Susceptibility measurements at high pressures using a microcoil system in an anvil cell. *Rev. Sci. Instrum.* **74**, 4728 (2003).

31. P. Panissod, G. Krill, C. Vettier, R. Madar, Antiferromagnetic metallic state of NiS2. *Solid State Commun.* **29**, 67–70 (1979).

32. Y. Feng, R. Jaramillo, A. Banerjee, J. M. Honig, T. F. Rosenbaum, Magnetism, structure, and charge correlation at a pressure-induced Mott-Hubbard insulator-metal transition. *Phys. Rev. B.* **83**, 35106 (2011).

33. T. Miyadai, S. Sudo, Y. Tazuke, N. Mori, Y. Miyako, Magnetic properties of pyrite type NiS2−xSex. *J. Magn. Magn. Mater.* **31-34**, 337–338 (1983).

34. K. Kikuchi, T. Miyadai, T. Fukui, H. Itô, K. Takizawa, Spin Structure and Magnetic Properties of NiS 2. *J. Phys. Soc. Japan*. **44**, 410–415 (1978).

35. C. Schuster, M. Gatti, A. Rubio, Electronic and magnetic properties of NiS2, NiSSe and NiSe2 by a combination of theoretical methods. *Eur. Phys. J. B*. **85**, 325 (2012).

36. P. Raybaud, J. Hafner, G. Kresse, H. Toulhoat, Ab initio density functional studies of transition-metal sulphides: II. Electronic structure. *J. Phys. Condens. Matter*. **9**, 11107–11140 (1997).

37. W. M. Temmerman, P. J. Durham, D. J. Vaughan, The electronic structures of the pyrite-type disulphides (MS2, where M = Mn, Fe, Co, Ni, Cu, Zn) and the bulk properties of pyrite from local density approximation (LDA) band structure calculations. *Phys. Chem. Miner.* **20** (1993), doi:10.1007/BF00208138.

38. J. Zaanen, G. Sawatzky, J. Allen, Band gaps and electronic structure of transition-metal compounds. *Phys. Rev. Lett.* **55**, 418–421 (1985).

39. S. Sudo, Metal-insulator transition and magnetic properties in the NiS2−xSex system. *J. Magn. Magn. Mater.* **114**, 57–69 (1992).

40. S. Miyasaka *et al.*, Metal-Insulator Transition and Itinerant Antiferromagnetism in NiS 2 - x Se x Pyrite. *J. Phys. Soc. Japan*. **69**, 3166–3169 (2000).

41. O. P. Welzel, F. M. Grosche, Patterned anvils for high pressure measurements at low temperature. *Rev. Sci. Instrum.* **82**, 033901 (2011).

42. V. Petricek, M. Dusek, L. Palatinus, Crystallographic Computing System JANA2006: General features. *Zeitschrift für Krist. - Cryst. Mater.* **229** (2014), doi:10.1515/zkri-2014-1737.

43. P. Kwizera, M. Dresselhaus, D. Adler, Electrical properties of NiS_{2-x}Se_{x}. *Phys. Rev. B*. **21**, 2328–2335 (1980).

44. P. Blaha, K. Schwarz, G. Madsen, D. Kvasnicka, J. Luitz, WIEN2k (2014), (available at http://www.wien2k.at/).

45. J. P. Perdew, K. Burke, M. Ernzerhof, Generalized Gradient Approximation Made Simple. *Phys. Rev. Lett.* **77**, 3865–3868 (1996).

46. A. Kokalj, XCrySDen—a new program for displaying crystalline structures and electron densities. *J. Mol. Graph. Model.* **17**, 176–179 (1999).

47. P. M. C. Rourke, S. R. Julian, Numerical extraction of de Haas - van Alphen frequencies from calculated band energies. *Comput. Phys. Commun.* **183**, 16 (2008).

48. R. C. Clark, J. S. Reid, The analytical calculation of absorption in multifaceted crystals. *Acta Crystallogr. Sect. A Found. Crystallogr.* **51**, 887–897 (1995).

49. P. J. Becker, P. Coppens, Extinction within the limit of validity of the Darwin transfer equations. I. General formalism for primary and secondary extinction and their applications to spherical crystals. *Acta Crystallogr. Sect. A*. **30**, 129–147 (1974).

**Acknowledgments:** The authors would like to thank G Lonzarich for fruitful discussions. This work is supported by the EPSRC through grant EP/K012894/1. SF acknowledges support by the ERC and the Alexander von Humboldt foundation. A portion of this work was performed at the National High Magnetic Field Laboratory, which is supported by NSF DMR-1157490 and the State of Florida. SWT, WAC, and DEG were supported in part by DOE NNSA SSAA DE-NA0001979.

**Author contributions:** Single crystals were grown and characterized by SF and MG. Resistivity measurements were conducted by SF. Quantum oscillations studies with the TDO were developed by HC, WAD, and FMG and carried out with SF, PR, XC, DG, and ST. High pressure methods were supported by PA. The manuscript was prepared by SF and FMG with the help of HC.

**Competing interests:** The authors declare that they have no competing financial interests.

**Data and materials availability:** Raw data will be made available from the Data repository at the University of Cambridge.

## H2: Figures and Tables

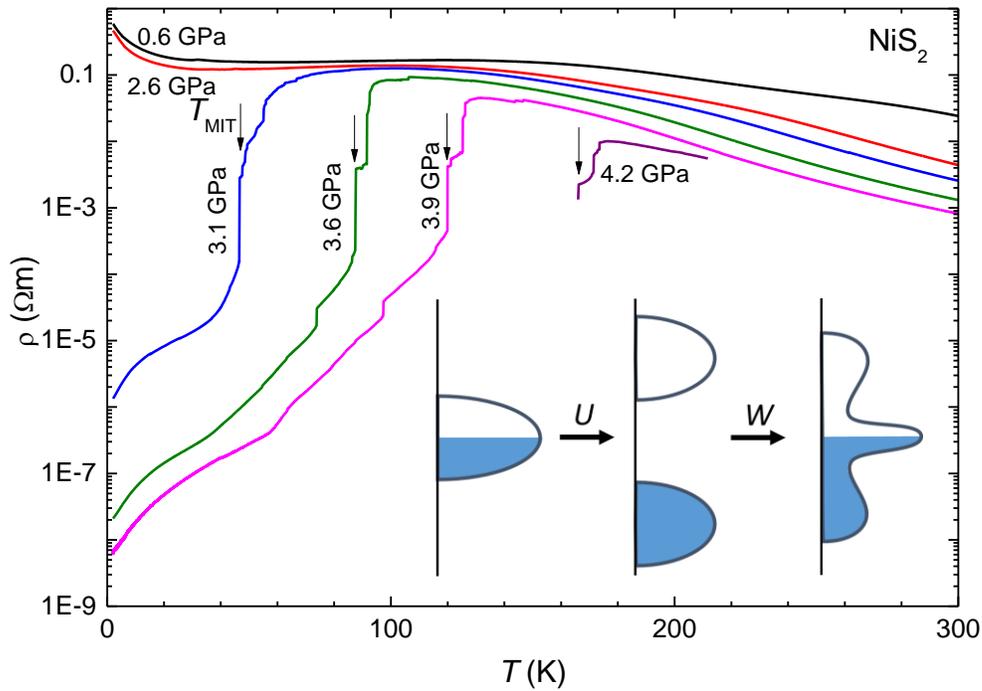

**Fig. 1. Resistivity of NiS$_2$ under pressure.** For pressures larger than 2.6 GPa, a rapid drop in the resistivity is observed at a temperature $T_{MIT}$, defined from the steepest slope. (inset) Schematic representation of the formation of upper and lower Hubbard bands for large Coulomb interaction $U$ (4), and of the emergence of a coherent quasiparticle peak at the chemical potential, when the bandwidth $W$ is increased, for instance by tuning the lattice density under pressure (5, 6).

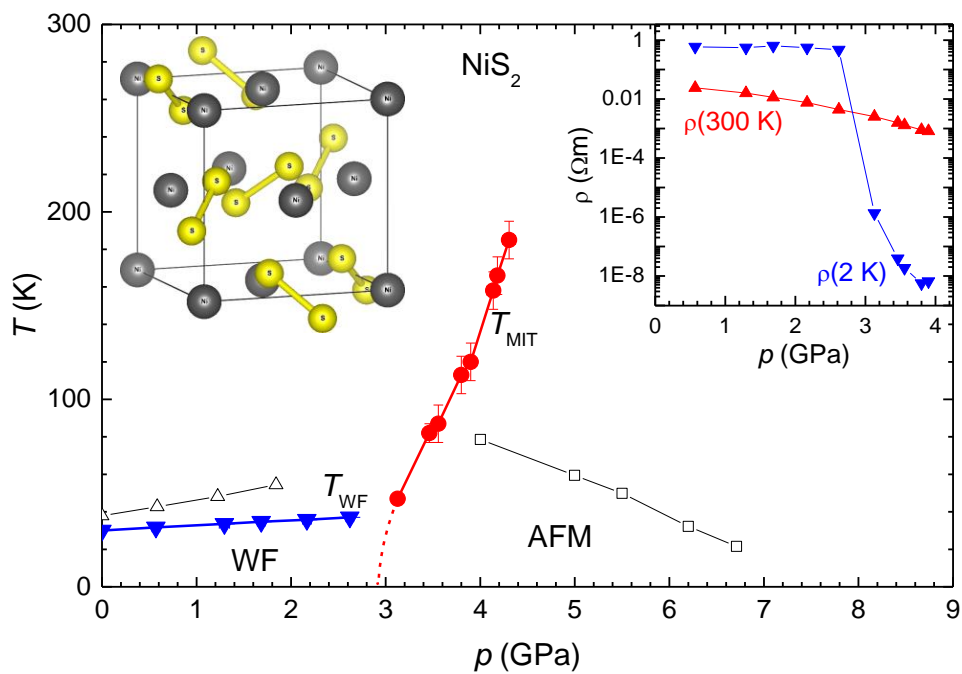

**Fig. 2. High-pressure phase diagram of NiS₂.** Solid circles and triangles represent the transition temperatures of the MIT and ferromagnetic order (cf. Supplementary information II) as extracted from resistivity, respectively. Open squares and triangles reflect the magnetic transition temperatures in the metallic and insulating state from Refs. (*29*) and (*28*), respectively. Left inset shows the pyrite structure of NiS₂ with the sulphur dimers indicated. Right inset shows the evolution of room-temperature (red triangles) and low-temperature (blue triangles) resistivity.

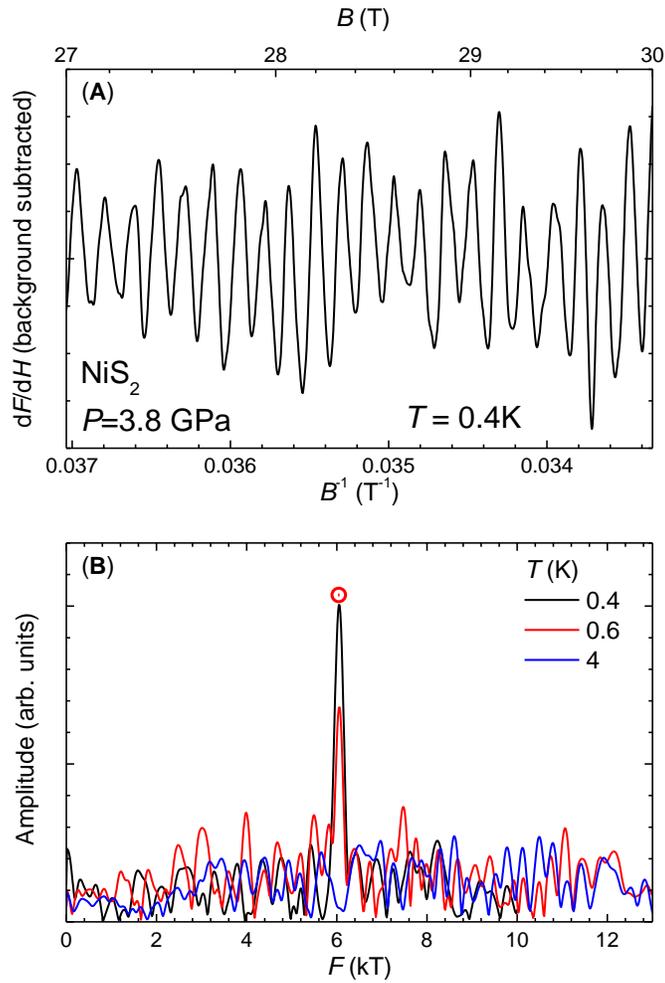

**Fig. 3. Quantum Oscillations in the metallic phase of NiS₂.** (**A**) We analyze the derivative d*F*/d*H* of the TDO frequency *F* with respect to field *H*. Numerical differentiation used locally fitted polynomials. A smooth background has been subtracted. Plotting against inverse magnetic field reveals the characteristic periodicity in 1/*H* of quantum oscillations. Data obtained from several sweeps with different sweep rates are averaged thus ruling out parasitic signals as a source of 1/*H* periodicity. (**B**) Power spectra of the Fourier transformed signal were obtained at several temperatures.

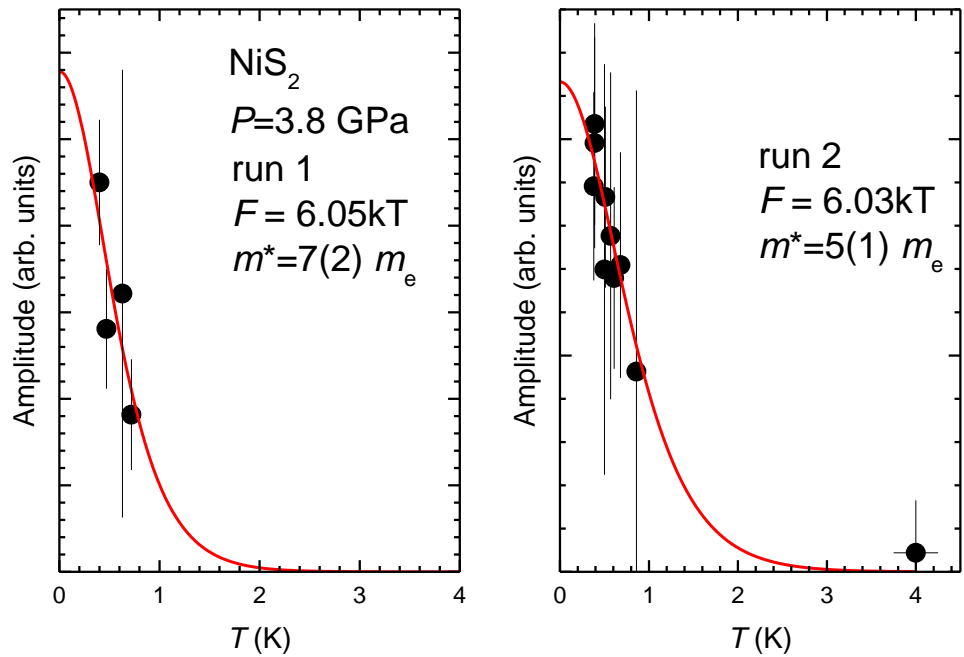

**Fig. 4. Determination of the effective mass.** The temperature dependence of the quantum oscillation amplitude (solid circles) is fitted with the Lifshitz-Kosevich form for two subsequent runs (solid red line). Vertical lines reflect standard errors estimated from background in the Fourier spectrum close to 6 kT (cf. Fig. 3).

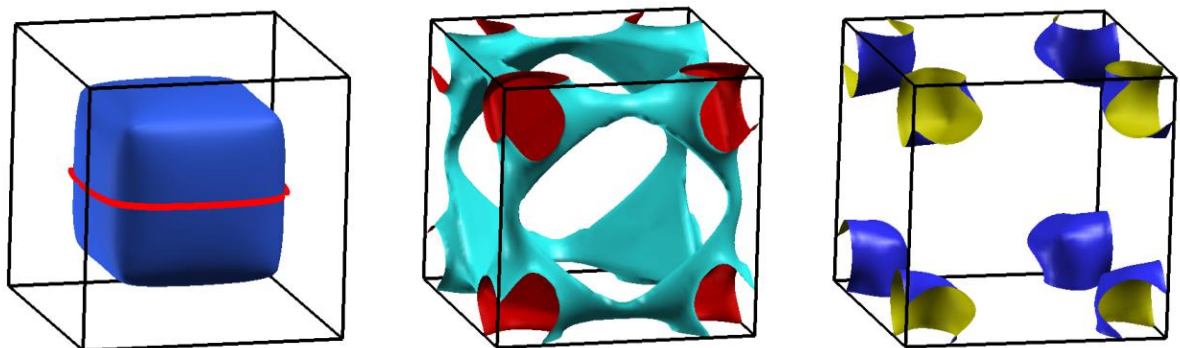

**Fig. 5. Calculated Fermi surface.** The Fermi surface was determined from band structure calculations using the lattice parameters and atomic positions as determined by our x-ray diffraction (see Supplementary information I). Besides the major sheets depicted here we find two small pockets in the Brillouin-zone corners. The solid red line on the first sheet represents its belly orbit.

| NiS$_2$ | Calc. | Exp. |
|---|---|---|
| $F$ (kT) | 6.3 | 6.03 |
| $m^*$ ($m_e$) | 0.8 | 6(2) |

**Table 1. Comparison of band structure calculation and experiment.** Frequency and effective mass calculated for the belly orbit of the "Cube" Fermi surface sheet are compared to those obtained from our QO measurements.

# H2: Supplementary Materials

## Supplementary Information I: Characterization of high-quality NiS$_2$ single crystals

Structural characterization of our crystals was carried out at room temperature using an Xcalibur E Single Crystal Diffractometer. Details concerning data collection and handling are summarized in table S1. Powder x-ray diffraction measurements were performed using a Bruker D2 Phaser diffractometer (Cu K-α radiation, 2$\theta$ interval of 3–80 degrees, step size of 0.018 degree, expose time of 4 hours) on powder prepared by grinding crystals with Si as an internal standard. Structure refinements based on both single crystal diffraction data and powder diffraction patterns were carried out using the Jana2006 program (*42*). X-ray diffraction measurements confirm that our crystals adopt the Pa-3 space group, the pyrite–type structure (*27*, *43*). The relevant crystallographic information is listed in table S1. Consistent results were obtained for several crystals across the batch. According to the single crystal X-ray diffraction study, our crystals are stoichiometric and have both Ni and S sites fully occupied (Table S2). Refinements with initial values of occupancies set as 0.95 converged with negligible populations of vacancies of less than ~0.1% for both crystallographic sites. This needs to be contrasted with previous studies on crystals grown using vapor transport technique, for which ~2% of S sites were found to be vacant.(*27*, *43*). Furthermore, our diffraction measurements performed on a number of crystals with typical dimensions of 0.1–0.2 mm showed mosaicities of 0.4–0.5 degrees. Such low values of the mosaicity parameter indicates a high degree of perfection of the crystal lattice.

The temperature-pressure phase diagram as constructed from our resistivity measurements is very similar to results published earlier (*28*, *29*). However, we find a slightly larger critical pressure of ~2.9 GPa compared to 2.5 GPa reported in earlier studies. The full phase diagram can be scaled between our studies and the earlier results on vapor-transport grown samples. This is illustrated in Fig. S1: Both the MIT and the magnetic transition coincide for our samples and the vapor-transport samples, when a scaling factor of 1.28 is applied to the pressure axis. This indicates a smaller compressibility of our crystals, which we attribute to the absence of vacancies.

## Supplementary Information II: Magnetic ground state in the insulating phase

A sizeable magnetization develops in our crystals below a transition temperature $T_{WF}$, suggesting a ferromagnetic component, in agreement with previous data. This transition can be interpreted as spin canting within an overall antiferromagnetic state and manifests in the resistivity as a kink, which we can follow under pressure in the insulating phase as demonstrated in Fig. S2(B). The pressure dependence as shown in Fig. S1 agrees with published data as described in Supplementary information I.

## Supplementary Information III: Band structure calculation

Band structure calculations were performed using the WIEN2k density functional code (*44*). Band energies were obtained on a 100000k-point mesh in the first Brillouin zone utilizing the generalized gradient approximation to the exchange correlation potential (*45*).

Relaxing the crystal parameters yielded good agreement with the experimentally observed values for both the lattice parameter and the sulphur position, as shown in table S2. This supports the validity of using the experimental structural parameters for our band structure calculations. We use the experimental lattice parameters determined for our sample and take into account a pressure-induced reduction of the lattice constant by 1.4% at 3.8 GPa. This reduction includes a correction for the reduced compressibility as discussed in Supplementary Information I over the reduction of 1.5% in Ref. (*32*). We find 5 bands crossing the Fermi energy with both nickel and sulphur character. The Fermi surfaces are visualized in Fig. 5 with the XCrySDen program (*46*). Extremal orbits and effective masses were extracted using the SKEAF algorithm capable of finding extended orbits across several reciprocal unit cells (*47*).

All orbits predicted for the orientation of the crystal used in our quantum oscillation experiment are listed in table S3. Only the belly orbit of the "Cube" is in agreement with the observed frequency of 6 kT, all other orbits are below 3 kT.

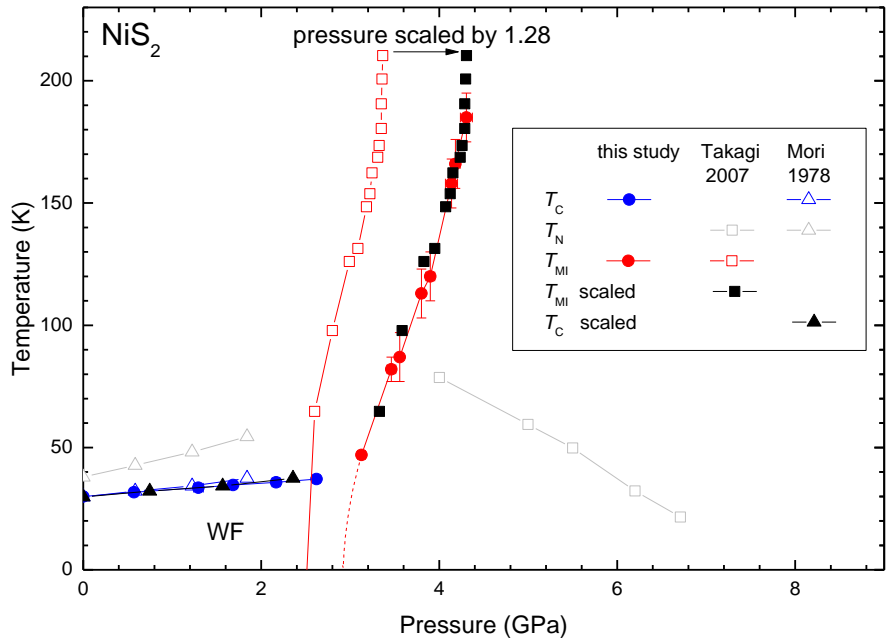

**Figure 1: Phase diagram of NiS$_2$ – Comparison with vapor-transport grown samples.** The MIT transition temperature $T_{MIT}$ and the transition temperature $T_{WF}$ into the weak ferromagnetic state (WF) of vapor-transport grown samples (*28*, *29*) can be scaled to our results using a factor of 1.28 for the pressure axis as illustrated with black solid squares ($T_{MIT}$) and triangles ($T_{WF}$).

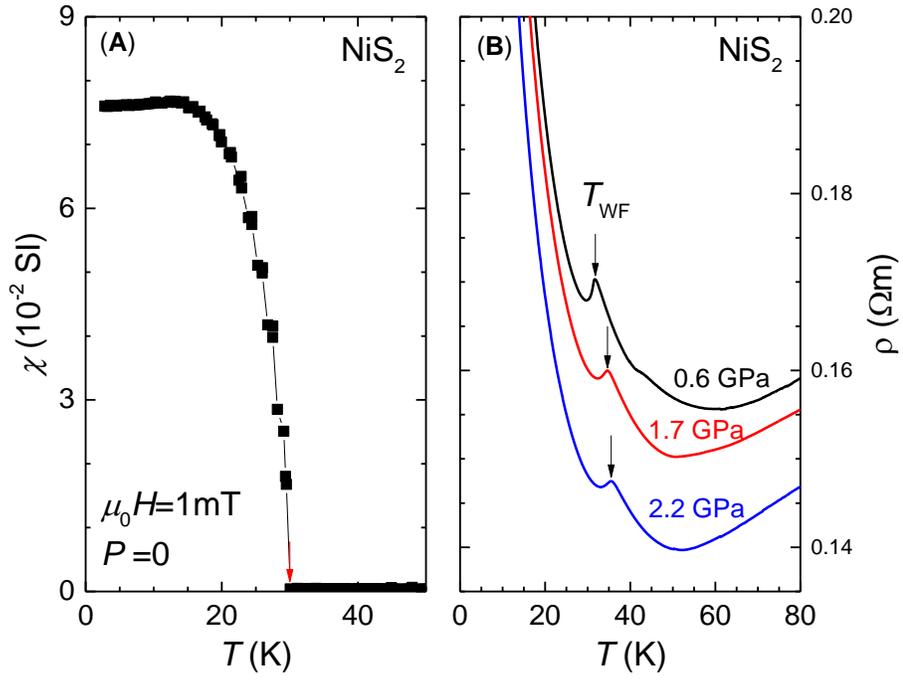

**Fig. S2. Ferromagnetic transition in the insulating phase.** (A) Ambient pressure magnetization measurements show the ferromagnetic ground state below $T_{WF}$ indicated by the arrow. (B) Resistivity measurements show the pressure dependency of $T_{WF}$ within the insulating phase.

| | |
|---|---|
| Structure type | pyrite |
| Space group | Pa-3 (No. 205) |
| Formula units/cell | 4 |
| Diffraction system | Xcalibur E, four-circle Kappa Sapphire CCD Detector (Xcalibur) |
| Radiation, $\lambda$ (Å) | Mo K$\alpha$, 0.71073 |
| Temperature (K) | 295(5) |
| Range in h,k,l | ±12, ±12, ±12 |
| R(eqv)/R($\sigma$) | 0.076/0.013 |
| $2\theta_{min}/2\theta_{max}$ | 0.076/0.013 |
| Observation criteria | F(hkl) > 3.00 $\sigma$(F) |
| Resolution d(Å) | 0.45 |
| Absorption coefficient | Face-based, analytical (48) |
| Absorption coefficient | 20.670 |
| N(hkl) measured | 23925 |
| N(hkl) unique | 364 |
| Extinction method | isotropic type 2 correction (49) |
| Extinction coefficient | 8200(300) |
| Goodness-of-fit (GOF) | 1.13 |
| R | 1.59% |
| wR | 1.83% |

**Table S1. Crystallographic data for NiS$_2$.**

| NiS$_2$ | | Te-flux | | | | | VT | calculation |
|---|---|---|---|---|---|---|---|---|
| Atom | Wyckoff site | $x$ | $B_{11}$ | $B_{13}$ | $B_{iso}$ | Occupancy | Occupancy | $x$ |
| Ni | 4a (0, 0, 0) | - | 0.517(4) | 0.011(2) | 0.517(2) | 1 | 1 | - |
| S | 8c (x, x, x) | 0.10531(2) | 0.464(4) | 0.017(2) | 0.464(2) | 1 | 0.98 | 0.10595 |
| a(Å) | | 5.6893(5)$^a$ | | | | | 5.689 | 5.698 |

$^a$powder data

**Table S2. Lattice parameters, atomic positional and displacement parameters for NiS$_2$** (Note: $B_{12}=B_{13}=B_{23}$ and $B_{11}=B_{22}=B_{33}$ for 4a site, $B_{13}=B_{23}=-B_{12}$ and $B_{11}=B_{22}=B_{33}$ for 8c site). The structural data on our Te-flux grown crystals are compared to those for crystals grown with vapor transport technique (VT) (*27, 43*).

| Band | Orbit | $F$ (kT) | $m^*$ ($m_e$) |
|---|---|---|---|
| 56 | 56a | 6.3 | 0.8 |
| 57 | 57a | 0.04 | 0.3 |
|  | 57b | 2.8 | 3.8 |
| 58 | 58a | 2.5 | 3.6 |
|  | 58b | 3.0 | 1.8 |
| 59 | 59a | 0.6 | 1.3 |
| 60 | 60a | 0.6 | 1.3 |
| exp |  | 6.03 | 6(2) |

**Table S3. Predicted frequencies on all Fermi surface sheets.**